\documentclass[conference]{IEEEtran}

\IEEEoverridecommandlockouts
\usepackage{cite}
\usepackage{amsmath,amssymb,amsfonts}
\usepackage{algorithmic}
\usepackage{graphicx}
\usepackage{textcomp}
\usepackage{xcolor}
\def\BibTeX{{\rm B\kern-.05em{\sc i\kern-.025em b}\kern-.08em
    T\kern-.1667em\lower.7ex\hbox{E}\kern-.125emX}}

\usepackage{setspace}
\begin{document}

\title{High-throughput Pairwise Alignment with the Wavefront Algorithm using Processing-in-Memory
\thanks{This work is supported by the University Research Board of the American University of Beirut (URB-AUB-104107-26306).}
}

\author{
    \IEEEauthorblockN{
        \vspace{-20pt}\\
        Safaa Diab\textsuperscript{1},
        Amir Nassereldine\textsuperscript{1},
        Mohammed Alser\textsuperscript{2},
        Juan G\'{o}mez Luna\textsuperscript{2},
        Onur Mutlu\textsuperscript{2},
        Izzat El Hajj\textsuperscript{1}
    }
    \IEEEauthorblockA{
        \textit{
        \textsuperscript{1}American University of Beirut, Lebanon~~~~~
        \textsuperscript{2}ETH Z\"{u}rich, Switzerland
        }
        \\\vspace{-20pt}
    }
}

\maketitle

\begin{abstract}

We show that the wavefront algorithm can achieve higher pairwise read alignment throughput on a UPMEM PIM system than on a server-grade multi-threaded CPU system.

\end{abstract}

\section{Introduction and Methods}
%\vspace{-2mm}

Pairwise sequence alignment is a fundamental computation in genome analysis.
The wavefront algorithm (WFA)~\cite{marco2021fast} is currently the state-of-the-art gap-affine pairwise alignment algorithm.
Aligning a massive number of genomic sequence pairs simultaneously on traditional CPU systems is inhibited by the memory bandwidth bottleneck.
Hence, sequence alignment is a memory-bound computation~\cite{lavenier2020variant}.

Processing-in-Memory (PIM) architectures~\cite{mutlu2019processing, mutlu2020modern} promise to alleviate the memory bandwidth bottleneck.
The UPMEM PIM architecture~\cite{devaux2019true} is the first PIM system to be commercialized in real hardware.
The hardware has been shown to be effective at accelerating memory-bound workloads~\cite{gomez2021benchmarking1,gomez2021benchmarking2}, including variant calling~\cite{lavenier2020variant}.
In this work, we introduce the first efficient PIM implementation of WFA for the UPMEM architecture.

A UPMEM DIMM is a standard DDR4-2400 DIMM consisting of multiple PIM chips.
Each PIM chip has multiple DRAM Processing Units (DPUs).
A DPU is a multi-threaded in-order 32-bit RISC core that can support up to 24 hardware threads.
Each DPU has exclusive access to a 64MB DRAM bank called Main RAM (MRAM) and a 64KB SRAM-based scratchpad memory called Working RAM (WRAM) which has lower latency than MRAM.
Programmers explicitly transfer data between MRAM and WRAM using DMA calls which must be 8-byte aligned, and perform load and store operations on WRAM.
Different DPUs cannot communicate with each other.
The CPU can communicate with the DPUs by transferring data to/from their MRAM banks.

In our PIM implementation, one CPU thread distributes read pairs evenly across DPU MRAMs using parallel data transfers.
Next, each DPU thread aligns multiple read pairs independently from other DPU threads to avoid the overhead of inter-thread synchronization.
A thread fetches the read pair from MRAM to WRAM, aligns the reads using WFA, and writes the result to MRAM.
When the DPUs complete, the CPU thread transfers the results back from the DPU MRAMs.

For fairness, we apply no optimizations to the WFA PIM implementation compared to the original WFA CPU implementation.
In fact, we remove vectorization from the PIM version because it is not supported on UPMEM.
The main challenge with implementing the PIM version is managing memory on the UPMEM architecture.
We replace WFA's original memory allocator with a custom allocator that manages the WRAM-MRAM hierarchy and overcomes its alignment restrictions.
Moreover, since a DPU's 64KB WRAM is shared among all threads, we cannot fit the WFA metadata for all threads in WRAM without sacrificing the number of threads.
Hence, to unleash the maximum threads, we store the metadata in MRAM and transfer it to/from WRAM on demand.

\section{Results}
%\vspace{-2mm}

\begin{figure}[h]
    \centering
    \vspace{-5pt}
    \includegraphics[width=\columnwidth]{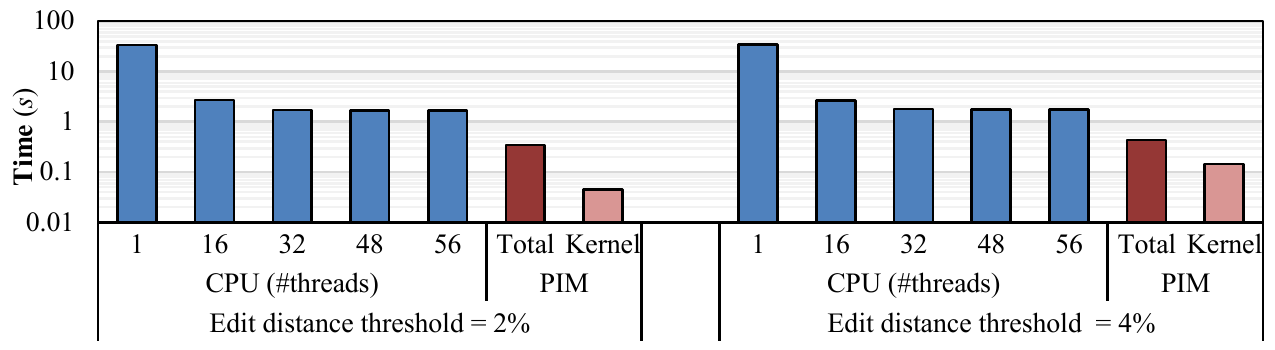}
    \vspace{-20pt}
    \caption{Time for aligning 5 million read pairs using WFA}\label{fig:results}
    \vspace{-5pt}
\end{figure}

Fig.~\ref{fig:results} compares the execution time of the original WFA implementation executed on a server-grade CPU with our proposed PIM implementation executed on a UPMEM system at full scale, when aligning 5 million pairs of 100bp-long reads with edit distance thresholds ($E$) of 2\% and 4\%.
The CPU system is dual socket with two Intel Xeon Gold 5120 processors (56 total threads) and 64 GB of memory.
The UPMEM system has 20 UPMEM DIMMs (2560 DPUs) clocked at 425MHz.

We make two key observations:
    (1) Performance does not scale well with the number of threads on the CPU system, which is expected since its performance is limited by memory bandwidth.
    (2) Our implementation (\texttt{Total}) has 4.87$\times$ and 4.05$\times$ higher throughput than the 56-thread CPU implementation for $E$ = 2\% and 4\%, respectively.
    The throughput is even higher (37.4$\times$ and 12.3$\times$, respectively) when the CPU-DPU data transfer time is not accounted for (\texttt{Kernel}).
Our future work includes scaling our implementation to longer read lengths and higher edit distance thresholds, and comparing to PIM implementations of other alignment algorithms.

\bibliographystyle{IEEEtran}
\bibliography{main}

\end{document}